# Use of group theory and Clifford algebra in the study of generalized Dirac equation for particles with arbitrary spin

I. I. Guseinov

*Department of Physics, Faculty of Art and Sciences, Onsekiz Mart University, Çanakkale, Turkey*

**Abstract**

Using condition of relativistic invariance, group theory and Clifford algebra the $2(2s+1)$-component Lorentz invariance generalized Dirac equation for a particle with arbitrary mass $m$ and spin $s$ is suggested, where $s = 0, \frac{1}{2}, 1, \frac{3}{2}, 2, ...$ In the case of half-integral spin particles, this equation is reduced to the $2s+1$ sets of two-component independent matrix equations. It is shown that the relativistic scalar and integral spin particles are described by $2(2s+1)$-component equation.

## 1. Introduction

The first higher spin equations have been proposed by Dirac in [1]. These equations in the presence of an external electromagnetic field, as was shown by Fierz and Pauli [2], led to the inconsistencies. They have suggested the equations for the special cases of $s = \frac{3}{2}$ and $s = 2$. Rarita and Schwinger [3] have developed theory of spin $\frac{3}{2}$ free particles which contains many of the features of the Dirac theory. The theory of spin-s free particles has been also developed by Proca [4], Kemmer [5] and Bargmann and Wigner [6]. All of these formalisms for spin-s free particles have many intrinsic contradictions and difficulties when an electromagnetic field interaction is introduced (see [7] and references therein). It should be noted that the mathematical structure of our study is different from all of the approaches which are available in the literature. Therefore, the generalized Dirac equation presented in this work can not be reduced to them. By the use of group theory and Cliffford algebra, we have shown in [8] that the generalized Dirac equation for the particles with arbitrary half-integral spin is consistent and causal in the presence of an electromagnetic field interaction. The aim of this work is, using the method set out in [8], to establish the generalized Dirac equation for particles with arbitrary spin. It should be noted that the arguments given for the solution of this problem are based on three completely different points of view, namely, the application of group theory, making use of the conditions of relativistic covariance and Lorentz invariance.

**2. Use of group theory and Clifford algebra**

For a single particle of charge $e$ and mass $m$ the relativistic Hamilton operator is given by

$$\hat{H} = \sqrt{c^2(\hat{\vec{p}} - \tfrac{e}{c}\vec{A})^2 + m^2c^4} + eA_0\,, \tag{1}$$

where $A_0$ is the scalar potential, $\vec{A}$ the vector potential and $\hat{\vec{p}} = \frac{\hbar}{i}\vec{\nabla}$ the momentum operator.

In accordance with the postulates of quantum mechanics the Hamilton operator $\hat{H}$ has to be linear and Hermitian. One can immediately see that the condition of linearity cannot be fulfilled, since the square root is not a linear operator. Therefore, the Dirac problem for arbitrary spin can be viewed in terms of a special polynomial algebra [9, 10]. In a previous work [8], for the linearization of the square root in the case of half-integral spin we have used the group theory and Clifford algebra. The generalized Dirac problem for arbitrary mass and spin can be solved in a similar way.

Using the method set out in [8], we obtain for the order of the generalized Dirac group the following relation:

$$g_s = 8(2s+1)^2\,, \tag{2}$$

where $s = 0, \frac{1}{2}, 1, \frac{3}{2}, 2, ...$ This group has $\frac{1}{2}g_s + 1$ classes, therefore, $\frac{1}{2}g_s + 1$ irreducible representations. The dimensions $n_i$ for these representations are determined by

$$\sum_{i=1}^{\frac{1}{2}g_s+1} n_i^2 = g_s\,, \tag{3}$$

where

$$n_i = \begin{cases} 1 & for \;\; 1 \le i \le \frac{1}{2} g_s & (4a) \\ 2(2s+1) & for \;\; i = \frac{1}{2} g_s + 1 & (4b) \end{cases}.$$

It is well known that the commutative representations do not satisfy the conditions of a Clifford algebra (see Ref.[10]). Consequently, the one-dimensional representations (4a) do not satisfy the conditions of Clifford algebra because they are commutative, therefore, only the $2(2s+1)$ dimensional irreducible representations (4b) can be used. The results are presented in Table 1.

## 3. Generalized Dirac equation

Making use of Table 1 obtained from the application of group theory and condition of relativistic covariance we introduce the following $2(2s+1) \times 2(2s+1)$ Hermitian and unitary matrices:

$$\vec{\alpha}^s = \begin{pmatrix} 0 & 0 & . & . & . & 0 & \vec{\sigma} \\ 0 & 0 & . & . & . & \vec{\sigma} & 0 \\ . & . & . & . & \vec{\sigma} & 0 & 0 \\ . & . & 0 & \vec{\sigma} & . & . & . \\ . & . & \vec{\sigma} & 0 & . & . & . \\ 0 & \vec{\sigma} & 0 & . & . & . & . \\ \vec{\sigma} & 0 & . & . & . & 0 & 0 \end{pmatrix} \tag{5}$$

and

$$\beta^s = \begin{cases} \begin{pmatrix} I^s & 0^s \\ 0^s & -I^s \end{pmatrix} & for \; s = 0, \frac{1}{2}, \frac{3}{2}, ... & (6) \\ \begin{pmatrix} 0^s & 0^s \\ 0^s & 0^s \end{pmatrix} & for \; s = 1, 2, 3, ... & (7) \end{cases}.$$

These matrices satisfy

$$\alpha_k^s \beta^s + \beta^s \alpha_k^s = 0 \tag{8}$$

$$\alpha_k^s \alpha_l^s + \alpha_l^s \alpha_k^s = 2\delta_{kl} I^s. \tag{9}$$

It should be noted that, in the case of integral values of s the matrices $\beta^s$ in the form $\beta^s = \begin{pmatrix} I^s & 0^s \\ 0^s & -I^s \end{pmatrix}$ do not satisfy the condition (8), i.e.,

$$\alpha_k^s \beta^s + \beta^s \alpha_k^s \neq 0 \qquad for\; s = 1, 2, 3, ... \tag{10}$$

Therefore, Eq. (7) correspondences to the case of integral spin particles. The generalized Dirac equation corresponding to the matrices (5), (6) and (7) is defined as

$$i\hbar \frac{\partial \Psi^s}{\partial t} = \hat{H}^s \Psi^s \tag{11}$$

$$\hat{H}^s = c\vec{\alpha}^s (\hat{\vec{p}} - \tfrac{e}{c}\vec{A}) + mc^2 \beta^s + eA_0 \tag{12}$$

$$\Psi^s = \begin{pmatrix} \phi^s \\ \tilde{\phi}^s \end{pmatrix}, \tag{13}$$

where for integral spin (s=0,1,2,…)

$$\phi^s = \begin{pmatrix} \varphi^{s0} \\ \varphi^{s1} \\ . \\ . \\ . \\ \varphi^{s,2s-1} \\ \varphi^{s,2s} \end{pmatrix} \tag{14a}$$

$$\tilde{\phi}^s = \begin{pmatrix} \tilde{\varphi}^{s,2s} \\ \tilde{\varphi}^{s,2s-1} \\ . \\ . \\ . \\ \tilde{\varphi}^{s1} \\ \tilde{\varphi}^{s0} \end{pmatrix}, \tag{14b}$$

for half-integral spin $\left( s=\frac{1}{2},\frac{3}{2},\frac{5}{2},... \right)$

$$\phi^{s} = \begin{pmatrix} \chi^{s0} \\ \chi^{s2} \\ . \\ . \\ . \\ \chi^{s,2s-3} \\ \chi^{s,2s-1} \end{pmatrix} \tag{15a}$$

$$\tilde{\phi}^{s} = \begin{pmatrix} \tilde{\chi}^{s,2s-1} \\ \tilde{\chi}^{s,2s-3} \\ . \\ . \\ . \\ \tilde{\chi}^{s2} \\ \tilde{\chi}^{s0} \end{pmatrix} \tag{15b}$$

Here, $\varphi^{s\lambda}$, $\tilde{\varphi}^{s\lambda}$ *and* $\chi^{s\lambda}, \tilde{\chi}^{s\lambda}$ are the single- and two-component matrices, respectively. The two-component matrices are defined as

$$\chi^{s\lambda} = \begin{pmatrix} u^{s\lambda} \\ u^{s,\lambda+1} \end{pmatrix} \tag{16a}$$

$$\tilde{\chi}^{s\lambda} = \begin{pmatrix} \tilde{u}^{s,\lambda+1} \\ \tilde{u}^{s\lambda} \end{pmatrix}, \tag{16b}$$

where $0 \le \lambda(1) \le 2s$ and $0 \le \lambda(2) \le 2s-1$ for s=0,1,2,… and $s=\frac{1}{2},\frac{3}{2},\frac{5}{2}...$, respectively.

By the use of procedure described in Dirac’s papers [11, 12] it is easy to show that the generalized Dirac equation (11) satisfies the condition of Lorentz invariance.

In the special case of scalar particles (s=0), the generalized Dirac equation (11) has the form:

$$i\hbar \frac{\partial \Psi^0}{\partial t} = \hat{H}^0 \Psi^0 \tag{17}$$

$$\hat{H}^0 = c\vec{\alpha}^0 (\hat{\vec{p}} - \tfrac{e}{c}\vec{A}) + mc^2 \beta^0 + eA_0 \tag{18}$$

$$\Psi^0 = \begin{pmatrix} \varphi^{00} \\ \tilde{\varphi}^{00} \end{pmatrix}, \tag{19}$$

where

$$\vec{\alpha}^0 = \vec{\sigma},\ \beta^0 = \sigma_3 = \begin{pmatrix} 1 & 0 \\ 0 & -1 \end{pmatrix}. \tag{20}$$

Here, $\vec{\sigma}$ is formed by the Pauli matrices $\sigma_1, \sigma_2, \sigma_3$ and

$$\vec{\sigma}\hat{\vec{p}} = \frac{\hbar}{i} \begin{pmatrix} \frac{\partial}{\partial z} & \frac{\partial}{\partial x} - i\frac{\partial}{\partial y} \\ \frac{\partial}{\partial x} + i\frac{\partial}{\partial y} & -\frac{\partial}{\partial z} \end{pmatrix}. \tag{21}$$

As can easily be seen that the Eq.(17) for relativistic scalar particles is a first-order differential equation, while the Klein-Cordon equation forms a second-order differential equation. Therefore, one has to arrive immediately at the conclusion that the Klein-Cordon equation does not meet the requirement of the condition of relativistic covariance, namely, the condition of linearity for a relativistic Hamiltonian. This equation only partially satisfies the postulates of (relativistic) quantum mechanics.

## 4. Conclusions

In this study, we have generalized the Dirac's $spin-1/2$ theory to a relativistic theory for particles with arbitrary integral and half-integral spin. It is shown that this theory has the following properties:

(1) The generalized Dirac matrices are irreducible and Clifford algebraic.

(2) The generalized Dirac wave functions and matrices possess the $2(2s+1)$ independent components.

(3) The generalized Dirac equation satisfies the condition of Lorentz invariance.

(4) The generalized Dirac equation of half-integral spin particles is reduced to the $(2s+1)$ sets of two-component matrix equations.

(5) The relativistic scalar particles $(s=0)$ satisfy the two-component Dirac equation.

(6) The integral spin particles $(s=1,2,3,...)$ satisfy the $2(2s+1)$ component generalized Dirac equation.

The generalized Dirac theory presented in this work can be used in the solution of different problems of describing particles with arbitrary spin within the framework of relativistic quantum mechanics.

**Table 1.** Summary of the generalized Dirac group properties for $0 \le s \le \frac{7}{2}$

| $s$ | Group Order | No. of Classes | No. of 1-D irred | No. of 2-D irred | No. of 4-D irred | No. of 6-D irred | No. of 8-D irred | No. of 10-D irred | No. of 12-D irred | No. of 14-D irred | No. of 16-D irred |
|---|---|---|---|---|---|---|---|---|---|---|---|
| 0 | 8 | 5 | 4 | 1 | 0 | 0 | 0 | 0 | 0 | 0 | 0 |
| $1/2$ | 32 | 17 | 16 | 0 | 1 | 0 | 0 | 0 | 0 | 0 | 0 |
| 1 | 72 | 37 | 36 | 0 | 0 | 1 | 0 | 0 | 0 | 0 | 0 |
| $3/2$ | 128 | 65 | 64 | 0 | 0 | 0 | 1 | 0 | 0 | 0 | 0 |
| 2 | 200 | 101 | 100 | 0 | 0 | 0 | 0 | 1 | 0 | 0 | 0 |
| $5/2$ | 288 | 145 | 144 | 0 | 0 | 0 | 0 | 0 | 1 | 0 | 0 |
| 3 | 392 | 197 | 196 | 0 | 0 | 0 | 0 | 0 | 0 | 1 | 0 |
| $7/2$ | 512 | 257 | 256 | 0 | 0 | 0 | 0 | 0 | 0 | 0 | 1 |

Note: irred-irreducible representations